\documentclass[prl,twocolumn]{revtex4}
\usepackage{amsmath,amssymb,mathrsfs}
\usepackage[dvips]{graphicx}
\usepackage{hyperref}

\def\scr{\mathscr}

 \def\SV{{\scr V}}

    \def\<{\langle}         \def\>{\rangle}

  \def\V0{{\mathbf 0}}

  \def\Bk{{\mathbf k}}

  \def\B0{{\mathbf 0}}
\def\Br{{\bf r}}

  \def\BR{{\mathbf R}}
\def\BS{{\mathbf S}} \def\BT{{\mathbf T}}

\def\be{\begin{equation}}       \def\ee{\end{equation}}
\def\bea{\begin{eqnarray}}      \def\eea{\end{eqnarray}}

\newcommand{\x}{\mathbf{r}}
\newcommand{\vk}{\mathbf{k}}

\newcommand{\R}{\mathbf{R}}

\newcommand{\q}{\mathbf{q}}

\begin{document}

\title{Orbital order in Mott insulators of spinless p-band fermions}

\author{Erhai Zhao and W. Vincent Liu}
\affiliation{Department of Physics and Astronomy, University of
Pittsburgh, Pittsburgh, Pennsylvania 15260, USA}

\begin{abstract}
A gas of strongly interacting spinless p-orbital fermionic atoms in
2D optical lattices is proposed and studied. Several interesting new
features are found. In the Mott limit on a square lattice, the gas is
found to be described effectively by an orbital exchange Hamiltonian 
equivalent to a pseudospin-1/2 XXZ model.  For a triangular,
honeycomb, or Kagome lattice, the orbital exchange is geometrically
frustrated and described by a new quantum $120^\circ$ model. We
determine the orbital ordering on the Kagome lattice, and show how
orbital wave fluctuations select ground states via the order by
disorder mechanism for the honeycomb lattice. We
discuss experimental signatures of various orbital ordering. 

\end{abstract}

\maketitle

The electron orbital degree of freedom plays an important role in
correlated quantum materials such as transition metal oxides
\cite{Tokura2000}. Many intriguing phases observed in experiments
are attributed to the coupling of electron $d$-orbitals,
typically in the $t_{2g}$ or $e_g$ manifold, to the electron
charge, spin, and/or the lattice degree of
freedom~\cite{Kugel:82,Khaliullin:05}.  Understanding
the intricate interplay between them remains a theoretical challenge.
It is thus desirable to study simpler systems where the orbital degree
of freedom is disentangled from others, namely ``plain vanilla"
orbital ordering in a Mott insulator.

Rapid advances in loading and controlling alkali atoms on the excited
bands of optical lattices
\cite{Esslinger:05,Browaeys:05,Bloch-M:07,Anderlini+Porto:07} have
made it possible to investigate orbital ordering of bosons and
fermions in new settings
\cite{Isacsson:05,Kuklov:06,Liu+Wu:06,Wu+Liu+Moore:06,Wu:07,Wu-Zhai:07pre,Wu-DasSarma:0712}.
Here we show that strongly interacting single species (spinless)
$p$-orbital fermions on optical lattices can realize Mott states
characterized by orbital-only models \cite{Brink:04}. We consider the
case where the degenerate $p$-orbitals of the optical potential well are
partially occupied while the fully occupied $s$-orbital acts as ``closed shell" and
remain inert at low energy scales.  Thus, these atomic Mott insulators
differ significantly from their solid state counterparts of $d$
electron systems in orbital symmetry.  We derive the low energy
effective orbital exchange Hamiltonian of $p$-band fermions in the
strong coupling regime and commensurate filling for simple 2D lattices
and determine their long range orbital ordering patterns.

Mott states of $p$-band fermions, described by orbital-only models,
represent new examples of correlated states in condensed matter.
Although we refer to the emergent low energy symmetry of the
degenerate $p$-orbital states as pseudo-spin symmetry, it is not an
internal symmetry like the true spin degree of freedom, but rather
intrinsically spatial. As we shall demonstrate, in general $p_{x,y}$
orbital states transform as the components of a spinor under space
rotation, in a manner reminiscent of the Lorentz spinor. This sets the
orbital exchange models apart from the familiar spin exchange models.

\paragraph{Spinless lattice $p$-band fermions.} We consider a
gas of fermion atoms in a single hyperfine spin state loaded in an optical
lattice interacting via a generic two-body potential
$\SV(\Br_1-\Br_2)$. Atoms interact predominately in the $p$-wave
channel, i.e., in momentum space, 
$\SV(\Bk^\prime-\Bk)\simeq 3\SV_1(k)(\hat{\Bk}^\prime\cdot\hat{\Bk})$.
The actual form of
$\SV_1(k)$ is unimportant for our discussion \cite{ho:090402}, as long
as it reproduces the low energy $p$-wave scattering amplitude
$f_1(k)=k^2/(-v_1^{-1}+c_1k^2-ik^3)$, where $v_1$ is the scattering
volume and $c_1$ is the effective range parameter
\cite{Gurarie:07}. These parameters are known for
$^{40}$K \cite{Ticknor+:multiplet:04}. For example, one can use a
separable model potential \cite{ho:090402} or a pseudopotential
\cite{Pricoupenko06} with coupling constant $g_p=12v_1\pi\hbar^2/m$
($m$ is the fermion mass). The strength of $\SV_1$ can be tuned using a
$p$-wave Feshbach resonance
\cite{Regal+Jin:p_wave:03,gunter:230401,zhang:030702,schunck:045601}. We
only consider repulsive interaction here.

We focus on a setup that captures the physics of the orbital symmetry and
quantum degeneracy of fermionic atoms. We first consider a
strongly anisotropic 3D cubic optical lattice potential,
$V_{op}(\x)=\sum_{\mu=x,y,z} V_\mu\sin^2(k_Lr_\mu)$ with $V_x=V_y$ and
$V_z\gg V_{x,y}$, where $k_L$ is the wavevector of the laser field and
we set the lattice spacing to be unit, $a=\pi/k_L\equiv1$. In the deep
lattice limit, $V_\mu\gg E_R\equiv \hbar^2k^2_L/2m$ (the recoil
energy) for all three directions, the bottom of the optical potential
at each lattice site can be approximated by a harmonic oscillator, and
the lowest few energy levels are $s, p_x, p_y, p_z$ orbital
states. The $|p_x\>$ and $|p_y\>$ orbitals are degenerate, while
$|s\>$ ($|p_z\>$) is well below (above) them, separated in energy by
the harmornic oscillator frequencies $\hbar\omega_\mu=\sqrt{4V_\mu
E_R}$. Due to the strong lattice potential in $z$-dimension, the
system is dynamically separated into a stack of approximately
independent layers with suppressed tunneling in between,
each being two-dimensional (2D).
Then, in order to explore the orbital-related quantum dynamics
in such a lattice, one can fill fermions up to the $p$-band by
having an average number of fermions per site between $1$ and $2$. In
the atomic limit, for two particles per site, one of them occupies the
$s$-orbital and the other one occupies either $p_x$ or $p_y$. We shall
refer to this as half filling.

We expand the fermion field operator in the Wannier basis,
$\psi(\x)=\sum_{i,\ell} w_\ell(\x-\R_i)c_{i\ell}$, where $\BR_i$ is
the lattice vector at site $i$ in two dimensions and
$\ell=s,p_x,p_y,p_z,...$ is the band
index. Then the interacting system is described by a multi-band
Hubbard model. Close to half filling, with the filled $s$ band and the
empty 
$p_z$ bands well separated from the degenerate $p_x$ and $p_y$ bands,
the low energy effective model for the fermions is the following 2D
$p$-band Hubbard Hamiltonian,
\be
H_p=\sum_{i;\mu,\nu=x,y}t_{\mu\nu}
[c^{\dagger}_{i,\mu}c_{i+\nu,\mu}+h.c.]
+ {U}\sum_i n_{ix}n_{iy}\,. \label{hubbard} \ee
Here,
$t_{\mu\nu}=t_{\parallel}\delta_{\mu\nu}+t_{\perp}(1-\delta_{\mu,\nu})$
is the hopping amplitude of orbital $\mu$ in the $\nu$ direction [the
orbitals will be denoted as $(p_x,p_y)\equiv (x,y)\equiv
(\uparrow,\downarrow)$ interchangeably to lighten the notation], and
$U$ is the onsite repulsion between atoms in $p_x$ and $p_y$ orbital
states. In the harmonic approximation of the Wannier function, the
transverse hopping $t_{\perp}=-e^{-(\eta/2)^2}V_x/2$ is negative and
small, while the longitudinal hopping $t_{\parallel}=(1-\eta^2
/2)t_{\perp}$ is positive and much larger in magnitude. Here,
$\alpha_\mu=(V_\mu/E_R)^{1/4}k_L$ is the inverse of the harmonic
oscillator length in the $\mu$-dimension, and $\eta\equiv\alpha_x a$ is typically a large number. The onsite energy
is given by $U=g_p\alpha_x^2\alpha_z(22\alpha_x^2+\alpha^2_z)/32(2\pi)^{3/2}$ in the pseudopotential approach.
Eq.\eqref{hubbard} looks like the ordinary one-band Hubbard model for spinless fermions, albeit with a twist: the $p_x$ (pseudo-spin $\uparrow$) atoms prefer
hopping in the $x$ direction while the $p_y$ ($\downarrow$) atoms prefer hopping in the $y$ direction. Such anisotropy has dramatic consequences in the strong coupling limit.

\begin{figure}
\includegraphics[width=2.2in]{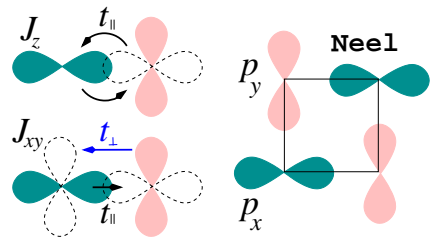}
\caption{(Color online) Left: Virtual hopping processes giving rise to the antiferro-orbital
Ising exchange $J_z$ and the ferro-orbital XY exchange $J_{xy}$. The transverse
hopping $t_{\perp}$ is much smaller than the longitudinal hopping $t_{\parallel}$ and has opposite sign.
Therefore $J_{xy}\ll J_z$. Dash lines indicate empty orbitals. Right:
Neel orbital ordering on the square lattice.}
\label{fig:hopping+Neel}
\end{figure}

\paragraph{Effective orbital exchange Hamiltonian.} At half filling and in
the strong coupling limit, $U\gg t_{\parallel}$, orbital fluctuation
is the only remaining low energy degree of freedom. It is well known that virtual hopping processes lead to direct and higher order orbital exchange
interactions \cite{Kugel:82}. In our case, as shown in
Fig.~\ref{fig:hopping+Neel},  the
nearest neighbor orbital exchange is strongly anisotropic as a direct
consequence of the anisotropic shape of the $p$-orbitals.
The effective Hamiltonian for $H_p$ can be derived following
the standard canonical transformation method. Up to order of
$t^2_{\parallel}$, it is given by $H^{(2)}=-U^{-1}T^{(2)}_{-1,1}$ in the
notation of Ref. \onlinecite{MacDonald:88}. We introduce pseudo-spin operator
$T^+=c^\dagger_xc_y$ and
$T^z= (c^\dagger_xc_x-c^\dagger_yc_y)/2$, or equivalently
$\mathbf{T}={1\over 2} c^\dagger_{\mu}\boldsymbol{\sigma}_{\mu\nu}c_{\nu}$ in the Cartesian vector form with $\boldsymbol{\sigma}$ being the standard Pauli matrices, and  rewrite $H^{(2)}$ as,
\be
H_{\rm orb}=\sum_{<i,j>}[J_{xy}(T^+_iT^-_j+h.c.)
+ J_{z} T_i^zT_j^z ]. \label{xxz}
\ee
Replacing $\BT$ with the usual spin $\BS$ in a magnetic system,
Eq.\eqref{xxz} corresponds to the familiar $XXZ$ model. The
antiferro-orbital Ising exchange
$J_{z}=2(t_{\perp}^2+t_{\parallel}^2)/U$ results from the {
longitudinal (and transverse) virtual exchange hopping.}  By contrast,
the $XY$ exchange $J_{xy}=2t_{\perp}t_{\parallel}/U$ involves both
longitudinal and transverse hopping, it is ferro-orbital ($J_{xy}<0$)
due to the opposite sign of $t_{\parallel}$ and $t_{\perp}$.  Because the
transverse hopping {amplitude} of orbitals is small, we have
$|J_{xy}|\ll |J_{z}|$, so the leading order interaction takes the form
of antiferro-orbital Ising model for pseudospin $T=\frac{1}{2}$,
\be
H_{\mathrm{orb}}\simeq J_{z} \sum_{<i,j>} T_i^zT_j^z.
\label{ising}
\ee
We neglect three body and ring exchange terms which are of higher
order in $t_{\parallel}/U$. The Ising exchange favors antiparallel
configuration of nearby pseudospins, i.e.  perpendicular configuration
of nearby $p$ orbitals (one in $p_x$ and the other $p_y$). Thus, the
Mott state is antiferro-orbitally ordered on square lattice. The
translational symmetry is broken with an alternative arrangement of
$p_x$ and $p_y$ orbitals as shown in Fig.~\ref{fig:hopping+Neel}.

\paragraph{Frustrated 120$^\circ$ model on oblique lattices.}
We now generalize the analysis to oblique 2D lattices.
{ For any vector $e_\theta$ directing at angle $\theta$ with the $x$ axis, such as the bond direction indicated in Fig.~\ref{fig:oblique},
it is convenient to introduce a local coordinate system with axis
$\tilde{x}$ and $\tilde{y}$ rotated from the global coordinate system
by $\theta$. The p-orbital wave functions transform under rotation
as the components of a planar vector, i.e.,
\be
\begin{array}{lcl}
c_x\rightarrow \tilde{c}_x &=& c_x\cos\theta +c_y\sin\theta\,, \\
c_y\rightarrow \tilde{c}_y &=& -c_x\sin\theta +c_y\cos\theta.
\end{array}
\label{eq:c->c}
\ee
Accordingly, the pseudospin operators transform as
\be
\begin{array}{lcl}
T_z &\rightarrow& \tilde{T}_z =T_x\sin 2\theta +T_z\cos 2\theta\,,\\
T_x &\rightarrow& \tilde{T}_x = T_x\cos 2\theta -T_z\sin 2\theta\,,\\
T_y &\rightarrow& \tilde{T}_y =T_y\,.
\end{array}\label{eq:T->T}
\ee
In other words, the pseudospin vector $\mathbf{T}$ is
rotated by $2\theta$ about the $y$ axis in the pseudospin space.
The leading order exchange interaction between site $i$ and $j$, connected by
bond $e_\theta$, is then given by $J_z \tilde{T}_i^z\tilde{T}_j^z$. Note the
value of longitudinal hopping along $e_\theta$, and consequently
$J_z$, for a specific lattice depends on details of the optical potential.
\begin{figure}
\includegraphics[width=2.6in]{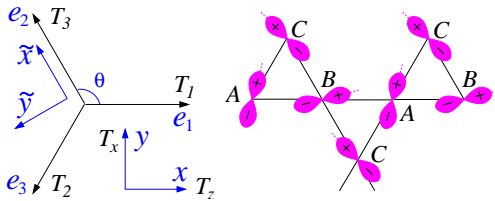}
\caption{(Color online) Left: The spatial coordinate system (in blue). Basis
vectors $e_j$ are at 120$^\circ$ from each other. Three pseudospins
$T_j$ (in black) defined on bond $e_j$. The quantization axis $T_z$ is
$x$ ($p_x$). Right: Orbital ordering on an individual triangle $ABC$
and long range order on the kagome lattice.}
\label{fig:oblique}
\end{figure}

We are particularly interested in the triangular and honeycomb
lattice, both of which can be  implemented using interfering
laser beams. For these lattices (lattice spacing $a=1$), we introduce
three basis vectors: $\hat{e}_1\equiv \hat{x}$, $\hat{e}_2$, and
$\hat{e}_3$, which are 120$^\circ$ from each other as shown in
Fig.~\ref{fig:oblique} and use $T_j$ to denote the operator
$\tilde{T}_z$ along} bond $e_j$ ($j=1,2,3$). Explicitly, $T_{1}=T_z$,
$T_{2}=-\frac{1}{2}T_z-\frac{\sqrt{3}}{2}T_x$, and
$T_{3}=-\frac{1}{2}T_z+\frac{\sqrt{3}}{2}T_x$, all lying within the $xz$
plane. Then the orbital exchange Hamiltonian can be put into a simple
form,
\be
H_{120}=J_z\sum_{\BR,j} T_{j}(\BR)T_{j}(\BR+\hat{e}_j), \label{h120}
\ee
where the sum over $\BR$ includes all lattice sites (or the sites
within the $A$ sublattice) for the triangular (or honeycomb)
lattice. We call Eq. (\ref{h120}) the quantum $120^{\circ}$ model, as
a formally similar Hamiltonian was proposed to describe the two-fold
degenerate $e_g$ electrons in cubic perovskites
\cite{Brink:04,Brink:99,Mostovoy:02,You:07}.  The underlying physics
is however very different. There, $e_j$  label the
cartesian basis vectors of the {\it cubic} lattice, while the
$120^{\circ}$ configuration of three $T_j$'s arises from a 
fundamentally different transformation property of two $d$-orbitals,
$|3z^2-r^2\rangle$ and 
$|x^2-y^2\rangle$, under the permutation of $x,y,z$.

The antiferro-orbital exchange favors perpendicular
alignments of nearest neighbor orbitals along bonds. Apparently, it is
impossible to achieve this for all three bonds on an elementary
triangle. Nor is it possible for three bonds joining at a site on the
honeycomb lattice. The orbital exchange is thus geometrically
frustrated on the triangular and honeycomb lattice. By frustration, we
mean the energy on each bond cannot be minimized simultaneously
\cite{Diep:04}. As we shall show below, the classical ground state
possesses a continuous degeneracy for the honeycomb lattice.  Solving the quantum
120$^\circ$ model is a nontrivial task, e.g., it remains an open
problem even for the cubic lattice
\cite{Brink:04,Brink:99,Mostovoy:02,You:07,Nussinov:04}. Here, in searching for the ground
state ordering pattern, we employ the semiclassical analysis which
proves fruitful in the study of frustrated magnets \cite{Murthy:97}: first
find the classical ground state, then consider the effect of leading
order orbital wave fluctuations.  Formally, this is done by generalize
$H_{120}$ to arbitrary psuedospin $T\geq \frac{1}{2}$ and consider the limit
of large $T$.

First we determine the classical ground state of $H_{120}$ on a
triangle cluster shown in Fig.~\ref{fig:oblique}. We parametrize the pseudospins at vertex $A,B,C$
with angles $\phi_{A,B,C}$ with respect to the $T_z$ axis ($AB$ bond). 
The classical energy, $E/T^2=f(0;A,B)+f(2\pi/3;A,C)+f(4\pi/3;B,C)$ where
$f(\gamma;A,B)\equiv\cos(\gamma-\phi_A)\cos(\gamma-\phi_B)$, is minimized when
$\phi_A-\phi_B=\phi_C-\phi_A=2\pi/3$, $\phi_A=5\pi/6$ or
$11\pi/6$. The two energy minima manifest the $C_2$ symmetry of $H_{120}$,
which is invariant under a  $\pi$-rotation of $\BT$ about the $y$
axis, i.e. $T_{x,z}\rightarrow -T_{x,z}$ with $\theta=\pi/2$ in
\eqref{eq:T->T}. 
The corresponding orbital configuration is shown in
Fig.~\ref{fig:oblique}, where 
the orbital at $A$ forms angle $\phi_A/{2}=75^\circ$ with the $AB$ bond.
Permutation of $(A,B,C)$ yields configurations of the
same energy. This result immediately tells us the classical orbital ordering
pattern (see Fig.~\ref{fig:oblique}) on the Kagome lattice, which consists of corner sharing
triangles.

\begin{figure}
\includegraphics[width=3.2in]{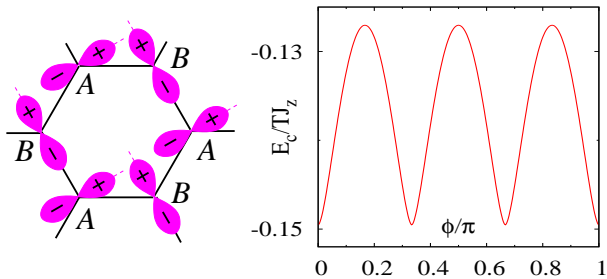}
\caption{(Color online) Left: Neel orbital ordering on the honeycomb 
lattice. The orbital at 
site $A$ forms an angle of $30^\circ$ with the horizontal $AB$ bond. 
Global rotation of all orbitals by $n\pi/6$ yields degenerate 
configurations. Right: The quantum fluctuation
correction to the ground state energy per unit cell, $E_c/TJ_z$,
which reaches minima at $\phi_n=n\pi/3$. This lifts the continuous degeneracy of the classical ground
states.}
\label{fig:honeycomb:order}
\end{figure}

\paragraph{Orbital order by disorder.}
For the bipartite honeycomb lattice, we introduce transformation
$T_{x,z}\rightarrow\bar{T}_{x,z}=\pm T_{x,z}$, with $+$ ($-$) for
sites within the $A$ ($B$) sublattice. Then the orbital exchange become
ferro-orbital in terms of
the new $\bar{\BT}$ operators, $H_{120}=$$-J_z\sum_{\BR\in A,j}
\bar{T}_{j}(\BR)\bar{T}_{j}(\BR+\hat{e}_j)$. It can be
rewritten (apart from a constant factor) as
\be
\bar{H}_{120}=\frac{J_z}{2}\sum_{\BR\in
A,j}[\bar{T}_{j}(\BR)-\bar{T}_{j}(\BR+\hat{e}_j)]^2\,, j=1,2,3\,.
\ee
Therefore, the classical energy is minimized for
any homogeneous configuration
$(\bar{T}^z,\bar{T}^x)(\BR)$=$(T\cos\phi,T\sin\phi)$,
independent of the polar angle $\phi$. This implies Neel (antiferro)
orbital ordering. Note that 
this continuous SO(2) degenerate manifold of classical 
ground states is not an inherent property of $H_{120}$
which is only invariant under finite point group rotations. Orbital
wave excitations are important quantum fluctuations beyond mean field theory.
To the leading order in $1/T$, their 
correction to the ground state energy 
can be computed following the spirit of Holstein-Primakov spin wave theory
\cite{Murthy:97}.
The algebra is cumbersome but straightforward.
The central result is the 
quantum correction to the ground state energy per unit cell (containing one 
$A$ site and one $B$ site),
\be
{E_c(\phi)\over T J_z}
=\frac{1}{N}\sum_{\Bk,\lambda}\omega_{\lambda}(\Bk)-\frac{3}{2}.
\ee
Here, index $\lambda=\pm 1$ labels the two branches of the 
orbital wave excitations with dispersion 
$\omega_{\lambda}(\Bk)=(\sqrt{3}/4)\sqrt{3+2\lambda
  |\beta_{\Bk}(\phi)|}$. 
The form factor 
$\beta_{\Bk}(\phi)=\sin^2(\phi)e^{i\Bk\cdot \hat{e}_1}+\sin^2(\phi-\pi/3)e^{i\Bk\cdot \hat{e}_2}+\sin^2(\phi+\pi/3)
e^{i\Bk\cdot \hat{e}_3}$. $N$ is the number of sites within the $A$ sublattice, and the $\Bk$ sum is within the
first Brillouin zone of the triangular sublattice. 
$E_c(\phi)$ is plotted in the right panel of Fig.~\ref{fig:honeycomb:order} for $\phi=0$ to $\pi$,
and $E_c(\phi+\pi)=E_(\phi)$. 
We see that orbital fluctuation lifts the SO(2) degeneracy and chooses ground 
state $\phi_n=n\pi/3$ ($n$ is any integer), where function $E_c(\phi)$ reaches minimum.
This mechanism is well known in frustrated spin systems as ``order by disorder"
\cite{Diep:04,Murthy:97}.  
The Left panel of Fig.~\ref{fig:honeycomb:order} shows a representative orbital
ordering pattern corresponding to $n=1$, i.e., the $p$-orbital at site $A$ is 
at angle $\phi_{n=1}/2=30^\circ$ with the horizontal $AB$ bond. Successive global rotation
of all orbitals by $\pi/6$ yields degenerate Neel configurations.

Finally we turn to the triangular lattice. We limit ourselves to
3-sublattice (R3) ordering \cite{Murthy:97} as one of the potential
orders for pseudospin $T=1/2$ ~\cite{Note:stripe}.  The R3 order is
parametrized by three polar angles, $\phi_{A,B,C}$, describing the
in-plane pseudospins at each sublattice $A,B,C$. The classical energy
per unit cell
$E/T^2=3[\cos(\epsilon_1)+\cos(\epsilon_2)+\cos(\epsilon_1+\epsilon_2)]/2$
only depends on two independent parameters $\epsilon_1=\phi_A-\phi_B$
and $\epsilon_2=\phi_B-\phi_C$, and it is minimized for
$\epsilon_1=\epsilon_2=\pm 2\pi/3$. Therefore, the classical R3
configuration is infinitely degenerate with its degeneracy
parametrized by $\phi_A$. Its orbital wave analysis will be presented
elsewhere.

To summarize, our mean-field and low energy fluctuation analysis suggests
new correlated Mott phases with long range $p$-orbital ordering
on the kagome, honeycomb, and triangular lattice. At finite
temperature, we expect thermal fluctuations stabilize orbital
ordering, as previously demonstrated for the cubic 120$^\circ$ model
\cite{Brink:04,Nussinov:04}. Our analysis is very suggestive but
cannot be taken as complete for pseudospin $T=1/2$.
There, quantum fluctuation may be strong enough to melt the long range orbital order,
raising the possibility of realizing a disordered ``orbital liquid"
ground state. Future work is needed to address this open question.

\paragraph{Detecting the orbital order.}  Here, we predict
experimental signatures of the orbital-ordered phases found above in the
density-density (noise) correlation function, $G(\Br,\Br';t)=\langle
n(\Br,t)n(\Br',t)\rangle-\langle n(\Br,t)\rangle\langle n(\Br',t)\rangle$,
which can measured in the time of flight image taken 
after time $t$.  We generalize the analysis of
Ref. \cite{Altman:04,AFHO:06,Liu+Wu:06} to the case of multi-orbital spinless
fermions and focus on various contributions from $\langle
c^\dagger_{\q,\mu} c_{\q,\nu}c^{\dagger}_{\q',\mu'}c_{\q',\nu'}\rangle$.  Here
$\q=m\Br/\hbar t+\mathbf{G}$ is the quasimomentum folded into the first
Brillouin zone of the underlying lattice, $\mathbf{G}$ is the
reciprocal wave vector, and $\mu,\nu$ are the orbital indices. The
closed-shell $s$-fermions contribute the usual anti-bunching dips at
$\mathbf{G}$, reflecting the Fermi statistics
\cite{Altman:04,Rom:06}. The $p$-fermions on the other hand contribute
additional terms such as structure factor $\langle
T^z_{\q}T^z_{\q'}\rangle$ and $\langle T^+_{\q}T^-_{\q'}\rangle$,
which lead to new dips at $\mathbf{G}_r$, the reciprocal lattice wave
vector of the enlarged unit cell in the ordered state.  For example,
$\mathbf{G}_r=(\pi,\pi)$, $(0,\pi)$,
$(2\pi/\sqrt{3},0)$, for the square, kagome, and 
honeycomb lattice, respectively.

\paragraph{Note Added.} After the submission of our manuscript, there appeared
Ref.~\cite{Wu:08pre} which independently proposed and studied 
a similar quantum $120^\circ$ model.

We thank C. Ho and V. Stojanovic for helpful discussions. This work is
supported in part by ARO W911NF-07-1-0293.

\bibliography{cold_atoms}

\begin{thebibliography}{35}
\bibitem[{Not()}]{Note:stripe}
\bibinfo{note}{An independent work of Wu \cite{Wu:08pre} suggests that the
  classical $H_{120}$ model develops strip order on the triangular lattice,
  which is in a different symmetry class than we considered.}
\end{thebibliography}
\bibliographystyle{apsrev}
\end{document}